# Charge and spin transport in nanoscopic structures with spin–orbit coupling

A. Reynoso, Gonzalo Usaj, C.A. Balseiro*

*Instituto Balseiro and Centro Atómico Bariloche, Comisión Nacional de Energía Atómica, 8400 San Carlos de Bariloche, Argentina*

## Abstract

During the last years there has been much interest, and theoretical discussion, about the possibility to use spin–orbit coupling to control the carriers spins in two-dimensional semiconducting heterostructures. Spin polarization at the sample edges may occur as the response of systems with strong SO-coupling to an external transport current, an effect known as spin Hall effect. Here, we show that in a 2DEG with Rashba SO-coupling, spin polarization near the sample edge can develop kinematically for low electron densities. We also discuss the effect in quantum wires where lateral confinement plays an important role.
© 2006 Elsevier B.V. All rights reserved.

## 1. Introduction

Semiconductor spintronics is a fast developing field that aims to build up new technologies for information processing based on the control and manipulation of the electronic spin degrees of freedom [1]. Systems with spin–orbit (SO) interaction are good candidates for *spintronics materials* as they provide a natural way to operate on the spin through the charge degree of freedom.

A large number of phenomena based on the SO interaction have been studied in semiconducting heterostructures, thin films and bulk materials [2]. In this work, we present a study of the effect of the SO coupling on the transport properties of nanoscopic systems tailored in two-dimensional electron gases (2DEG). In the case of 2DEG in semiconducting heterostructures, the asymmetric confining potential leads to a Rashba SO coupling [3]. The magnitude of the SO coupling strongly depends on the system, being small in n-doped GaAs–AlGaAs heterostructures and moderate or large in p-doped GaAs–AlGaAs or in InSb or $In_xGa_{1-x}As$-based heterostructures [2].

Among the remarkable effects predicted to occur in these systems it is worth mentioning the *spin-Hall effect* [4–7] and the *spin polarization of currents* in point contacts [8,9]. In particular, the former has attracted much attention during the last years. In analogy with the conventional Hall effect, the spin Hall effect refers to spin accumulation (magnetization) at the sample edges as a result of a transport current. The spin Hall effect may be due to impurities with SO coupling that produce a spin dependent scattering [5]. Recently, another mechanism has been proposed by Sinova et al. [6]. Such mechanism, which leads to an *intrinsic* spin Hall effect, may occur in 2DEGs. The idea is that in high mobility 2DEGs with Rashba coupling, a transport electric current generates a transverse spin current. However, it was argued that in a stationary state the zero frequency transverse spin current vanishes [10].

Experiments in semiconducting heterostructures [11], in quantum wells [12] and in GaAs thin films [13,14] observed a spin Hall effect. The optically detected magnetization at the edge of a Hall bar shows the presence of a spin polarization induced by a longitudinal electric current. Furthermore, in the case of 2DEG confined in a AlGaAs quantum well, the edge magnetization presents a complex structure with spacial oscillations. Although in thin films the effect has been attributed to an extrinsic—impurity scattering—mechanism [15] its origin in 2DEG confined in semiconducting heterostructures and quantum wells is still unclear.

*Corresponding author. Tel.: +59 2944 445261; fax: +54 2944 445299.
E-mail address: balseiro@cab.cnea.gov.ar (C.A. Balseiro).

The experimental observation of the spin Hall effect and the intense debate on its physical origin triggered a number of studies including numerical simulations and geometrical effects in nanoscopic systems [16–20]. In what follows, we present a study of the spin Hall effect in small systems emphasizing the effect of edge corrugation and sample geometry. In particular, we study the effect of a side cavity and of constrictions in the sample. The latter are of particular relevance for two reasons: (i) constrictions or quantum point contacts (QPC) are used as charge injectors or detectors [21–23] (ii) SO coupling can polarize currents passing through point contacts [8,9]. A device generating spin polarized currents based on QPCs that avoids the use of high magnetic fields or magnetic-semiconducting interfaces could become of central importance for spintronics. It is then interesting to analyze the current induced magnetization close to a constriction, the spin polarization of the transmitted current and the interplay between the two phenomena. The constriction can describe a QPC if it is short or a narrow quantum wire (QW) when it is long [24].

## 2. The model

We consider a 2DEG in the $(x, y)$ plane with Rashba SO interaction. The Hamiltonian of the system reads

$$H = \frac{p_x^2 + p_y^2}{2m^*} + \frac{\alpha}{\hbar}(p_y\sigma_x - p_x\sigma_y) + V(\mathbf{r}), \quad (1)$$

where $m^*$ is the effective mass, $\alpha$ is the Rashba coupling parameter and $V(\mathbf{r})$ is the confining potential in the plane of the 2DEG (lateral confinement). At the *bulk* of the 2DEG, the SO-coupling acts as an effective magnetic field contained in the 2DEG plane and with a magnitude that is proportional to the momentum of the carrier. This effective field lifts the spin degeneracy of the bands. At the edge of the sample, the confining potential $V(\mathbf{r})$ scatters the electrons changing its momentum. The effective field then rotates, generating a torque on the spin, an effect that could lead to different effects in confined systems.

In order to describe the system, we integrated the Hamiltonian numerically using a finite difference scheme that is equivalent to work with a tight-binding model [25,26]. The resulting model Hamiltonian reads

$$H = \sum_{n\sigma} \varepsilon_n c_{n\sigma}^\dagger c_{n\sigma} - t \sum_{n,\delta,\sigma} c_{n\sigma}^\dagger c_{n+\delta\sigma} + \text{h.c.}$$
$$- \lambda \sum_n \Big(\mathrm{i}\, c_{n\uparrow}^\dagger c_{(n+\hat{y})\downarrow} + \mathrm{i}\, c_{n\downarrow}^\dagger c_{(n+\hat{y})\uparrow}$$
$$- c_{n\uparrow}^\dagger c_{(n+\hat{x})\downarrow} + c_{n\downarrow}^\dagger c_{(n+\hat{x})\uparrow}\Big) + \text{h.c.}, \quad (2)$$

where $c_{n\sigma}^\dagger$ creates an electron at site $n$ with spin $\sigma_z = \sigma$ and energy $\varepsilon_n = 4t + V(\mathbf{r}_n)$, $t = \hbar^2/2m^* a_0^2$ for neighboring sites, $a_0$ is the lattice parameter, and $\lambda = \alpha/2a_0$. The summation is carried out on a square lattice where the coordinate of site $n$ is $\mathbf{r}_n = n_x\hat{x} + n_y\hat{y}$ with $\hat{x}$ and $\hat{y}$ the unit lattice vectors in the $x$ and $y$ directions, respectively, and $\delta = \hat{x}, \hat{y}$.

The conductance and spin Hall response of a ballistic system can be calculated by attaching ideal leads to the sample and using the Keldysh formalism. In the linear response regime the conductance is given by [27]

$$G = \frac{e^2}{\hbar}\mathrm{Tr}[\mathbf{\Gamma}^R \mathbf{G}^r(E_F)\mathbf{\Gamma}^L \mathbf{G}^a(E_F)] \quad (3)$$

and the current induced spin polarization in the three directions ($\eta = x, y$ and $z$) is [18,28]

$$\langle S_\eta(\mathbf{r}_n)\rangle = \frac{\hbar eV}{4}$$
$$\times \mathrm{Tr}[\sigma_\eta \{\mathbf{G}^r(E_F)(\mathbf{\Gamma}^L - \mathbf{\Gamma}^R)\mathbf{G}^a(E_F)\}_{n,n}]. \quad (4)$$

Here, the trace is taken on the spin variables, $\sigma_\eta$ is a Pauli matrix, $\mathbf{G}_{i,j}^r(\omega)$ and $\mathbf{G}_{i,j}^a(\omega)$ are the retarded and advanced $2 \times 2$ matrix propagators from site $j$ to site $i$ and the matrix elements are defined by the spin indices. $\mathbf{\Gamma}^{L(R)} = \mathrm{i}(\mathbf{\Sigma}_{L(R)}^r - \mathbf{\Sigma}_{L(R)}^a)$ and $\mathbf{\Sigma}_{L(R)}^{r(a)}$ is the retarded (advanced) self-energy due to the left (right) contact and $E_F$ is the Fermi energy.

## 3. Spin Hall effect

Here, we adopt the usual terminology: spin Hall effect stands for a current induced magnetization at the sample edge, perpendicular to the plane containing the carriers. We have shown that in clean 2DEGs, the interplay between Rashba coupling, sample edges and transport currents generates this effect [17]. As a plane wave is deflected by the edge of the sample, the spin rotates leading to an out of plane component. The sign of the out of plane spin component depends on the direction of motion. In the presence of a transport current, there is a preferred direction of motion and the spin polarization is unbalanced, leading to a net magnetization at the sample's edge. This current induced magnetization is large only for low electron densities. As the density increases, fast $2k_F$ oscillations mask and partially suppress the effect [28].

The simplest geometry that shows the effect is a Hall bar of width $L_y = N_y a_0$, which is defined by $V(\mathbf{r}_n) = 0$ for $1 \leqslant n_y \leqslant N_y$ and infinite otherwise. At the ideal leads the SO interaction is zero and is turned on at the lead-sample interface as shown in Fig. 1. A bar with $N_y$ sites in the transverse direction has $N_y$ transverse modes, each one contribute with a 1D-like van Hove singularity to the density of states and a contribution of $2e^2/h$ to the conductance. Both, the fine structure of the density of states and of the conductance observed in Fig. 1 are due to the scattering at the lead-sample interfaces. A map of the current induced spin polarization $\langle S_z\rangle$ in the $(y, E_F)$ plane is shown in Fig. 1c. Away from the van Hove singularities, the induced magnetization $\langle S_z\rangle$ is a smooth function of the Fermi energy and oscillates in the transverse direction. At the sample edges there is net magnetization. Opposite edges have a magnetization pointing in opposite directions (see also Fig. 2). Similar oscillating out of plane magnetization $\langle S_z \rangle$ at the edge of a 2DEG confined in a GaAs quantum well have been observed in Ref. [12]. In this configuration, the current also induces some in plane magnetization and away from the lead-sample interface we obtain $\langle S_x \rangle = 0$ and $\langle S_y \rangle \neq 0$ (not shown) as expected for an infinite 2DEG [17,28]. In fact, about 15 years ago, Edelstein showed that in a 2DEG with Rashba coupling, an external current generates some spin polarization contained in the gas plane and perpendicular to the current [29]. In the ideal Hall bar, the magnetization $\langle S_y \rangle$ is reminiscent of this effect although it is subject to finite size effects.

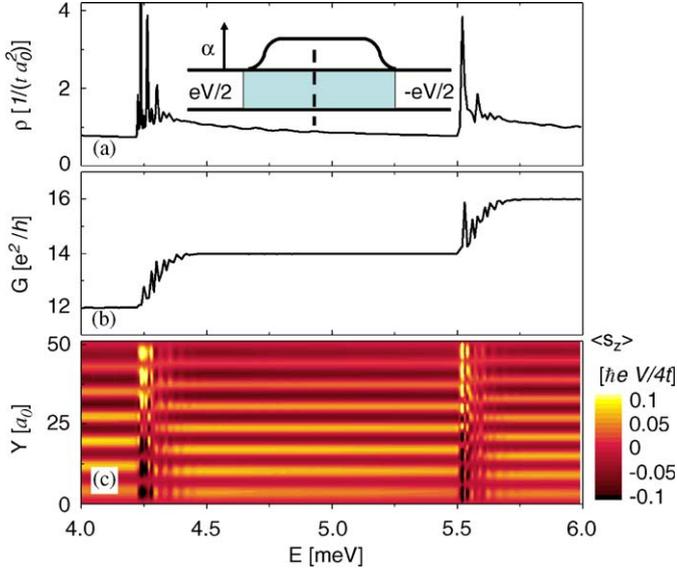

Fig. 1. (a) Local density of states at $x = 260a_0$ for a sample with $L_x = 500a_0$ and $L_y = 50a_0$ with $a_0 = 5$ nm. The inset shows a scheme of the sample with the spacial dependence of the SO coupling $\alpha$. (b) Conductance as a function of the Fermi energy. (c) Color map of $\langle S_z \rangle$ in the $(y, E_F)$ plane (for $x = 260a_0$). Here, we used $\alpha = 10$ meV nm and $m^* = 0.067m_0$.

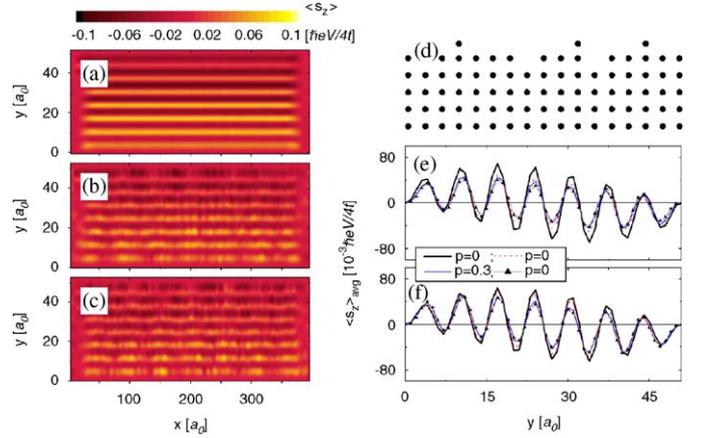

Fig. 2. Color map of $\langle S_z \rangle$ in the $(x, y)$ plane with (a) $p = 0$, (b) $p = 0.3$ and (c) $p = 0.5$. Parameters as in Fig. 1 and $E_F = 5$ meV. (d) Scheme of a corrugated edge. (f) Average magnetization $\langle S_z \rangle$ as a function of $y$ for different values of $p$ (as indicated) and $E_F = 5$ meV. (e) same as (f) but for $E_F = 5.3$ meV.

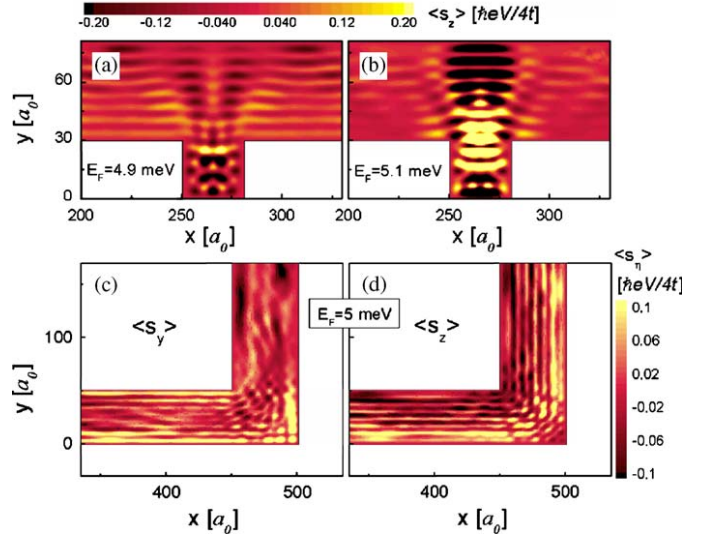

Fig. 3. (a) Color map of $\langle S_z \rangle$ in the $(x, y)$ plane. The total length of the system is $L_x = 530a_0$ and the other parameters as in Fig. 1. The size of the lateral structure is $30 \times 30a_0$. The Fermi energy is set to $E_F = 4.9$ meV in a) and to $E_F = 5.1$ in (b). The latter corresponds to a resonant state. (c) and (d) show the color map of $\langle S_y \rangle$ and $\langle S_z \rangle$ for a L-shaped sample.

How robust are these results against disorder or edge corrugation? To partially answer this question we have studied a simple model for edge corrugation. The corrugation is characterized by a parameter $p$ that gives the probability to find a defect a the edge: with equal probabilities $p/2$ we either remove or add a side site as shown schematically in Fig. 2. This procedure creates some corrugation with amplitude $2a_0$ without changing the mean width of the sample. The characteristic corrugation wavelength is $2a_0/p$. For each sample we calculate $\langle S_z \rangle$ and then make sample average typically with 10–20 samples. The results are shown in Fig. 2. For this model, our results show that the effect is not very sensitive to the edge disorder. Figs. 2e and 2f show the average magnetization across the bar. Increasing $p$ only produces a moderate reduction of the effect at the center on the sample.

More interesting is the effect of large structures. We have analyzed T-shaped and L-shaped structures. In the first case, shown in Fig. 3, the bar has a lateral structure. This structure is a cavity large enough to accommodate some low energy electronic resonances. The current induced magnetization in the structure strongly depends on the position of these resonances relative to the Fermi energy. Figs. 3a and b show the results for two values of the Fermi energy chosen to be away from any resonance and at a resonance, respectively. The local magnetization is quite large in the latter case. Changes in the local magnetization could also be obtained with a fixed Fermi energy by changing the resonance energy with a lateral gate potential.

The last example of this section is an L-shaped sample. In Figs. 3c and d, the electrons flow from left to right in the horizontal arm. The in-plane induced magnetization is perpendicular to the current so $\langle S_y \rangle \neq 0$ in the horizontal arm, as the electron current rotates to flow upwards in the vertical arm, $\langle S_y \rangle$ decreases and $\langle S_x \rangle$ becomes non-zero (not shown). Similar results were experimentally observed in L-shaped samples [14]. Conversely, the out of plane magnetization $\langle S_z \rangle$ has the same structure in both arms and away from the corner it reproduces the results of the straight sample.

## 4. Spin polarized currents

Despite of the fact that in 2DEG the current induces some in-plane magnetization, the current itself is not spin polarized [29]. The same occurs in some nanoscopic systems like the Hall bar described in the previous section. Notably, some nanoscopic structures like constrictions may generate polarized currents. This effect was recently discussed by Eto et al. [8] in the context of QPCs. Here we analyze the current though a QW, its spin polarization and the current induced magnetization close to the constriction.

A QW is described by including a contribution $V_C(\mathbf{r}_n)$ in the potential $V(\mathbf{r}_n)$. This potential is controlled by a parameter $V_g$ representing a gate voltage. We use the model for $V_C(\mathbf{r}_n)$ described in Ref. [27] (shown in Fig. 4). To analyze the spin polarization of the current, the conductance $G$ in Eq. (3) is separated in two contributions $G_+$ and $G_-$ describing the currents leaving the sample due to electrons with spin pointing in the $+y$ or $-y$ directions, respectively. The electrons transmitted through the QW are scattered by $V_C$ producing transitions between subbands of different spins. As a result, the spin of the transmitted flux is unbalanced ($G_+ \neq G_-$). Eto et al. [8] argued that the effect is controlled by the smoothness of the potential $V_C(\mathbf{r}_n)$. For a large gate voltage $V_g$, the QW is closed and the conductance is exponentially small. As $V_g$ is reduced the total conductance $G = G_+ + G_-$ increases and shows the $2e^2/h$ steps. In the first plateaus, the partial conductances $G_+$ and $G_-$ are different from each other. In the first one while $G_+ < e^2/h$, $G_- > e^2/h$ (see Fig. 4a). The fact that $G_-$ is larger than $e^2/h$ indicates that the two spin channels contribute. In fact, the two spin components propagate inside the QW. As the QW opens towards the wide region, the $+$ spin flips to the $-$ direction. Since the transition occurs in the region where the QW widens—where there are already more than one channel at $E_F$—the transition is not inhibited by the Pauli exclusion principle. In smooth potentials $V_C$ there is more *room* for these transitions as some channels penetrate in a region where $V_C$ is still effective to flip the spin. The spin polarization of the current, defined as $P = (G_+ - G_-)/(G_+ + G_-)$, is shown also in Fig. 4a. Its absolute value is almost one in the first plateau and decreases to smaller values in the higher plateaus. Note it also decreases in the tunneling region ($G < 2e^2/h$).

The structure of the current induced magnetization is shown in Figs. 4c and d. For gate voltages $V_g$ corresponding to the first conductance plateaus, the induced out of plane magnetization has its largest value in the region of the point contact. There $\langle S_z \rangle$ has a node along the QW axis and different signs at each side. The number of nodes increases with the number of channels contributing to the conductance. This is shown in Fig. 4d where $V_g$ is set to correspond to the second conductance plateau.

## 5. Summary and conclusions

We have analyzed the current induced polarization in nanoscopic systems. For 2D long bars, we obtained out of plane magnetizations that oscillate in the transverse direction. Close to each edge the induced magnetization $\langle S_z \rangle$ has a dominant sign. This result is a consequence of the interplay between the Rashba coupling, the confining potential and the external current that unbalances the number of carriers traveling in each direction. Using a simple model for edge corrugation, we showed that the results are not very sensitive to edge disorder.

Large structures, like a lateral cavity or a constriction may locally amplify the effect. In particular, narrow QWs create large out of plane magnetization and at the same time polarize the current. Spin polarized current are due to the spin-dependent scattering induced by the constriction. The spins are flipped as electrons go in and out of the constriction. This does not harm the out of plane magnetization inside the contact that has the same structure as in wider wires without a constriction.

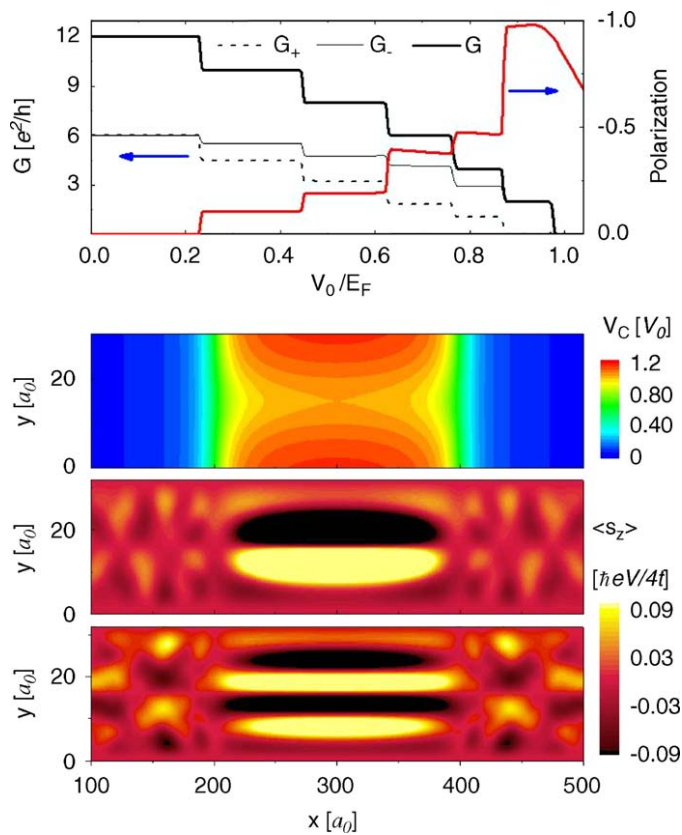

Fig. 4. (a) Total conductance, $G$, spin-resolved conductances, $G_-$ (thin solid line) and $G_+$ (dashed line), and polarization $P$ as a function of the gate voltage (here $V_0 \propto V_g$). (b) Potential landscape of the constriction, $V_C(\mathbf{r}_n)$. The current induced magnetization $\langle S_z \rangle$ corresponding to the first (second) plateau is shown in (c), (d). Here, $\alpha = 20$ meV nm and $E_F = 10$ meV.

One of the problems for the experimental study of spin polarized currents is how to detect them. We have shown that lateral electron focusing in system with strong SO coupling is an appropriate tool for a quantitative analysis of this effect [26]. In recent experiments Rokhinson et al. measured [30], using electron focusing in a hole-doped GaAs/AlGaAs structure, what seems to be spin polarized currents due to point contacts.

## Acknowledgments

We acknowledge partial support from CONICET, ANPCYT Grants no 03-13829 and 03-13476 and Fundación Antorchas: *Physics at the Nanoscale* and Grant 14169/21.